**Exchange-biased topological transverse thermoelectric effects in a Kagome ferrimagnet**

Heda Zhang[1,#], Jahyun Koo[2,#], Chunqiang Xu[1,3,#], Milos Sretenovic[1], Binghai Yan[2], and Xianglin Ke[1*]

[1]Department of Physics and Astronomy, Michigan State University, East Lansing, Michigan 48824-2320, USA

[2] Department of Condensed Matter Physics, Weizmann Institute of Science, Rehovot, Israel

[3] School of Physics, Southeast University, Nanjing 211189, China

#: These authors contributed equally to this work.

* Corresponding author: kexiangl@msu.edu

**Abstract**

**Kagome metal TbMn$_6$Sn$_6$ was recently discovered to be a ferrimagnetic topological Dirac material by scanning tunneling microscopy/spectroscopy measurements. Here, we report the observation of large anomalous Nernst effect and anomalous thermal Hall effect in this compound. The anomalous transverse transport is consistent with the Berry curvature contribution from the massive Dirac gaps in the 3D momentum space as demonstrated by our first-principles calculations. Furthermore, the transverse thermoelectric transport exhibits asymmetry with respect to the applied magnetic field, i.e., an exchange-bias behavior. Together, these features place TbMn$_6$Sn$_6$ as a promising system for the outstanding thermoelectric performance based on anomalous Nernst effect.**

**Introduction**

The study of topological materials with nontrivial electronic band topology has led to the prediction and realization of various topological insulators, metals, and semimetals [1-3]. In particular, magnetic topological materials are currently under intense research efforts owing to a variety of intriguing phenomena stemming from the integration of magnetism and topological electronic properties, such as quantized anomalous Hall effect [4], axion insulator [5], topological Fermi arc and chiral anomaly [6], etc. Notable examples of single-phase, intrinsic magnetic topological materials include the antiferromagnetic Weyl semimetal in $Mn_3(Sn/Ge)$ [7,8] and GdPtBi [9,10], the antiferromagnetic topological insulator in $MnBi_2Te_4$ [11,12], the ferromagnetic Weyl semimetal in $Co_3Sn_2S_2$ [13] and $Co_2Mn(Ga, Al)$ [14,15], the ferromagnetic Dirac metal in $Fe_3Sn_2$ [16-18].

Most of previous studies of magnetic topological materials have been focused on the non-trivial band topology, the corresponding surface states, and the intrinsic anomalous Hall effect (AHE) [19]. Charge carriers acquire an 'anomalous velocity' arising from the Berry curvature of electronic bands, which leads to the anomalous transverse conductivities. Recently, there has been surging interest in the anomalous Nernst effect (ANE) in magnetic topological materials due to the following two merits. From the scientific viewpoint, the anomalous Nernst coefficient is more sensitive than the AHE to the Berry curvature near the Fermi level, serving as a complementary tool in characterizing the Berry phase distribution in the reciprocal space [20,21]. From the viewpoint of technological applications, the ANE-based design allows for building malleable thermoelectric modules by patterning the thermopile onto flexible substrates [22-28]. This is illustrated by the schematic shown in Fig. 1(a), where magnetic topological materials with alternating magnetization are patterned next to each other and electrically connected in series, which can lead to a large thermoelectric voltage output. However, as the thermopile density continues to increase, the inherent magnetic stray field generated by neighboring modules may perturb the magnetization of individual modules, reducing the total output voltage [22]. One way to circumvent this issue is to pin the magnetization of modules, which can be achieved via the exchange-bias mechanism. Exchange-bias refers to a phenomenon in which the field-dependent hysteresis loops of a measured quantity shift along the applied magnetic field axis, which has been the most commonly observed in the magnetization data of heterostructures consisting of both ferromagnet and antiferromagnet [29]. More recently, exchange-bias phenomenon has also been observed in crystalline single-phase compounds [30,31]. Therefore, identifying materials hosting both large ANE as well as an exchange-bias feature is an important next step in realizing such designs.

Here we report comprehensive electrical, thermal, and thermoelectric measurements on a ferrimagnetic Kagome metal $TbMn_6Sn_6$. In addition to AHE, we show that $TbMn_6Sn_6$ exhibits both large ANE (2.2 $\mu V/K$ at 300 K) and anomalous thermal Hall effect (ATHE, 0.12 $W/m\,K$ at 300 K). Our first-principles calculations demonstrate that the large Berry curvature of the massive Dirac gaps in the 3D momentum space leads to the intrinsic anomalous transverse conductivities. Furthermore, we observe exchange-bias behavior in the field dependence of all these traverse transport quantities. This study highlights the potential of utilizing this magnetic topological material as a thermoelectric module.

**Results**

$TbMn_6Sn_6$ belongs to a member of the $RMn_6Sn_6$ (R = rare earth, Y, Sc, Lu) series. The manganese atoms in these compounds form a double-layer Kagome lattice structure, as illustrated in Fig. 1(b). Spins within each Kagome plane are ferromagnetically aligned while the spin alignment between Kagome planes depends on R elements [32,33]. The unique Kagome lattice can give rise to flat bands and Dirac points in the band structure [34,35]. The former provides an ideal platform for studying strongly correlated phenomena, while the latter is an interesting topological object by its own right. In particular, TbMn$_6$Sn$_6$ stands out as a ferrimagnet with out-of-plane magnetization where Tb moment aligns antiferromagnetically with Mn moment [36]. As a result, the electronic band structure of TbMn$_6$Sn$_6$ exhibits spin polarized Dirac dispersion with a Chern gap, as reported in Ref. [37,38] recently. The massive Dirac bands near the Fermi energy can give rise to anomalous electric and heat transport phenomena.

We first present the magnetic susceptibility measurements of TbMn$_6$Sn$_6$ in Fig. 1(c). The red solid curve and blue dashed curve shown here represent the field-cooled (FC) and zero-field-cooled (ZFC) susceptibility respectively measured along the crystalline c-axis with 0.1 T magnetic field, and the black solid curve represents the susceptibility measured along the a-axis. As seen in the inset of Fig. 1(c), there is a substantial decrease of $\chi_c$ above $T_{sr} \approx 309$ K, which is accompanied with a sharp increase of $\chi_a$. This is due to a spin re-orientation process as supported by neutron diffraction measurements [39]. Below $T_{sr}$, both Tb and Mn magnetic moments, which are antiferromagnetically coupled to each other, switch from an in-plane configuration to an out-of-plane structure. Note that TbMn$_6$Sn$_6$ undergoes paramagnetic-to-ferrimagnetic phase transition at $T_c = 423$ K [36]. Interestingly, the FC and ZFC $\chi_c$ ($T$) curves bifurcate noticeably below 200 K, which is presumably associated with the presence of slow magnetic fluctuation, as revealed by μSR recently, that leads to the formation of magnetic domains during the ZFC process [39]. As will be discussed later, such domain formation also gives rise to distinct hysteresis features in different temperature regions.

Figure 1(d) plots the temperature dependence of longitudinal transport coefficients ($\kappa_{xx}, \rho_{xx}$) of TbMn$_6$Sn$_6$. The resistivity, represented by the blue curve, decreases monotonically with temperature, and the residual resistance ratio is found to be 102.9, indicating good crystal quality. The thermal conductivity, shown in red, exhibits a typical broad peak around 30 K, resulting from the competition between boundary/defect scattering (at low temperature) and the Umklapp scattering (at high temperature). The Seebeck coefficient $S_{xx}$ as shown in Supplementary Fig. 1 in the Supplementary Information, linearly decreases with temperature above 160 K followed by a nearly constant feature in the temperature range between 80 K to 160 K. Phonon or magnon drag process may account for the slow decrease of $S_{xx}$ at low temperature. A similar behavior in Seebeck coefficient has been observed recently in Weyl semimetals Co$_2$MnGa and Co$_3$Sn$_2$S$_2$ [23,40].

Next, we present the field dependence of transverse transport coefficients of TbMn$_6$Sn$_6$. The AHE with an intrinsic anomalous Hall conductivity ($\sigma_{xy}^{int} = 0.14\ e^2/h$ per kagome bilayer) has been observed in TbMn$_6$Sn$_6$, which was attributed to the Berry curvature in Dirac bands [37]. For completeness, in Figure 2 we have included our own field-dependent anomalous Hall conductivity ($\sigma_{xy}^A$) data, together with the anomalous Nernst signal ($S_{xy}^A/T$) and anomalous thermal Hall conductivity ($\kappa_{xy}^A$). The sign conversion of transverse transport measurements and the subtraction of the normal transverse components of these quantities have been described in detail in the Supplementary Information (Supplementary Fig. 2, and Supplementary Fig. 3 –6). And all these three quantities measured at various temperatures are intentionally shifted along the Y-axis by a certain value for clarity. Note that the thermoelectric and thermal transport measurements were done in an adiabatic manner (Supplementary Fig. 7). There are several interesting features worth pointing out. First, the curves of $\sigma_{xy}^A$, $\kappa_{xy}^A$ and $S_{xy}^A$ measured at the same temperature have nearly identical features, implying that all these anomalous transverse transport properties stem from the same origin, i.e., the Berry curvature of the electronic bands. This is supported by the theoretical calculations as will be discussed latter. Second, the hysteresis loops measured above 200 K exhibit very small remanence, in contrast to the hysteresis loops measured at lower temperatures. The low remanence in hysteresis loops above 200 K is presumably due to the strong magnetic fluctuations, which is recently revealed by μSR measurements [39]. In contrast, magnetic fluctuations slow down at lower temperatures. Third, interestingly, an exchange-bias feature (i.e., the center of the hysteresis loop is shifted away from the origin along the field-axis) is observed in the hysteresis loops of $\sigma_{xy}^A$, $\kappa_{xy}^A$ and $S_{xy}^A$ measured below 200 K. This exchange-bias behavior will be discussed in further details later. The exchange-biased ANE observed in TbMn$_6$Sn$_6$ renders it as a promising material for realizing the proposed thermoelectric devices utilizing ANE.

In Fig. 2(d-e) we summarize the temperature dependence of the anomalous transport coefficients. The blue symbols in Fig. 2(d) represent $\sigma_{xy}^A(T)$ which increases monotonically with the decrease of temperature. The increase of $\sigma_{xy}^A$ at low temperatures mainly arises from the enhanced skew scattering ($\sigma_{xy}^{sk} \propto \sigma_{xx}$) contributions. It is unlikely that the side jump ($\sigma_{xy}^{sj}$) mechanism has a major contribution since the electric conductivity ($\sigma_{xx}$) of TbMn$_6$Sn$_6$ is high and therefore far away from the 'Localized-hopping conduction region' [19]. To extract the intrinsic anomalous Hall conductivity $\sigma_{xy}^{int}$, we assume the total AHE signal is composed of the intrinsic contribution and the extrinsic skew scattering contribution, i.e., $\rho_{yx}^A = \rho_{yx}^{sk} + \rho_{yx}^{int} = A\rho_{xx} + B\rho_{xx}^2$. By plotting $\rho_{xy}^A$ as a function of $\rho_{xx}$ (Supplementary Fig. 8) and fitting it to the equation above, the intrinsic anomalous Hall conductivity is extracted to be $\sigma_{xy}^{int} = 131 \pm 20 \ (\Omega \ cm)^{-1}$, which is in good agreement with the recent study.[37] The extracted $\sigma_{xy}^{int}$ is smaller than the measured $\sigma_{xy}^A$, suggesting the extrinsic contribution to $\sigma_{xy}^A$. The red symbols in Fig. 2(d) represent the measured anomalous thermal Hall conductivity normalized to temperature $(\kappa_{xy}^A/T)$, which shows a similar behavior to that of $\sigma_{xy}^A(T)$. Note that the scales of the left and right axes of Fig. 2(d) are related by the Lorenz number ($L_0 = 2.44 \times 10^{-8}\ V^2K^{-2}$). Alternatively, a plot of $L/L_0$ ($L = \frac{\kappa_{xy}^A}{\sigma_{xy}^A T}$) as a function of temperature is shown in Supplementary Fig. 9. One can clearly see that the 'anomalous' Wiedemann–Franz law ($\kappa_{xy}^A = L_0 T \sigma_{xy}^A$) [41] is obeyed for the anomalous transverse transport in TbMn$_6$Sn$_6$ within the measured temperature range. Figure 2(e) presents the temperature dependence of the anomalous Nernst coefficient ($S_{xy}^A$). While $S_{xy}^A$ monotonically decreases with temperature, it appears that $S_{xy}^A$ decreases in a faster manner in the higher temperature region compared to that at lower temperatures. Note that no anomalous Nernst signal is convincingly observed below 50 K.

To further our understanding of the origin of the anomalous Hall and thermoelectric conductivity in TbMn$_6$Sn$_6$, we performed density-functional theory (DFT) calculations within the generalized gradient approximation (GGA) framework [42,43]. Based on the calculated electronic structure, we projected the Bloch wave functions into Wannier functions to construct an effective Hamiltonian and evaluate the Berry curvature distributions in the reciprocal space. Figure 3(a) shows the calculated band structure and corresponding projected Berry curvature ($-\Omega_{xy}$). In all calculations, spin-orbit coupling (SOC) has been taken into account. The SOC opens a bandgap at Dirac points (K-points) near the Fermi energy E$_F$ [see an expanded view in Supplementary Fig. 10], giving rise to large Berry curvature with opposite signs for the upper and lower bands. In addition to the massive Dirac points, there are extra Berry curvature contributions for the band anti-crossing along $k_z$. Figure 3(b) presents the Berry curvature distribution in 3D Brillouin zone at E$_F$. Our results show that TbMn$_6$Sn$_6$ is a multi-band system beyond the simple kagome model. The massive Dirac points together with other bands contribute to the AHE.

Using the Kubo-formula approach [20], we calculated the intrinsic value of $\sigma_{xy}^{int}$ and the intrinsic transverse thermoelectric conductivity $\alpha_{xy}^{int}$. Figure 3(c) shows the calculated $\sigma_{xy}^{int}$ as a function of energy, where the Fermi energy $E_F$ is shifted down by 90 meV from the charge neutral point to optimally fit AHE and ANE. At the Fermi level, the calculated $\sigma_{xy}^{int} = 120 \ (\Omega \ cm)^{-1}$, which is in close agreement with the value extracted from the experimental results [Supplementary Fig. 8]. Figure 3(d) plots the comparison between the calculated anomalous thermoelectric conductivity normalized to temperature and the experimental results ($\alpha_{xy}^A/T$, blue dots), the latter of which are obtained using $\alpha_{xy}^A = \sigma_{xy}^A S_{xx} + \sigma_{xx} S_{xy}^A$ [44]. The yellow curve shows the temperature dependence of $\alpha_{xy}^{int}/T$ calculated using Kubo formula, $\alpha_{xy}^{int}(T) = -\frac{e}{\hbar} \int d\zeta \frac{\partial f(\zeta-\mu)}{\partial \zeta} \frac{\zeta-\mu}{T} \int_{BZ} \frac{d\vec{k}}{(2\pi)^3} \sum_{\epsilon_n<\zeta} \Omega_{xy}^z(\vec{k})$, wherein $f(\zeta-\mu) = \left(e^{\frac{\zeta-\mu}{k_B T}} + 1\right)$ is the Fermi-Dirac distribution function and $\mu$ is the chemical potential. We can see that while the calculated $\alpha_{xy}^{int}/T$ and the experimental values agree reasonably well at high temperature, they deviate at low temperature with the calculated values smaller than the experimental results. Such difference is mainly attributed to the enhanced skew scattering contribution to $\alpha_{xy}^A$ (via $\sigma_{xy}^A$) as the system approaches a 'clean limit' (increase of $\sigma_{xx}$). Similar features have been observed recently in other topological materials [14,22,23,40]. To take into account the effect due to the enhanced skew scattering, we recall the Mott relation ($\frac{\alpha_{xy}^A}{T}|_{T\to 0} = -\frac{\pi^2 k_B^2}{3|e|} \frac{d\sigma_{xy}^A}{d\zeta}|_\mu$), which states that the magnitude of $\alpha_{xy}^A/T$ is proportional to the derivative of $\sigma_{xy}^A$ relative to energy at the Fermi level. As mentioned previously, in addition to $\sigma_{xy}^{int}$, $\sigma_{xy}^A$ also includes the skew scattering contribution which can lead to ANE via the Mott relation. Thus, it is reasonable for us to scale the calculated $\alpha_{xy}^{int}/T$ value by a temperature dependent factor $\beta(T) = \sigma_{xy}^A/\sigma_{xy}^{int}$, which is shown by the orange curve in Fig. 3(d).

Although this is a coarse estimate of the skew scattering contribution, we can see that indeed the scaled, calculated $\beta\alpha_{xy}^{int}/T$ and the experiment values tend to merge at low temperatures. This feature, together with the closeness between the calculated $\alpha_{xy}^{int}/T$ and the experimental $\alpha_{xy}^{A}/T$ at high temperature, indicates the effects of band structure topology on the transverse thermoelectric transport properties in TbMn$_6$Sn$_6$.

We now discuss the asymmetric field dependence feature observed in the transverse transport measurements shown in Fig. 2(a-c). Such a feature is also observed in isothermal magnetization measurements conducted under the same experimental procedure, as shown in Supplementary Fig. 11, which indicates that magnetism, electronic, and thermal transport properties intimately correlate with each. To further affirm that the asymmetric field dependence feature is intrinsic and associated with the exchange-bias, we have studied both temperature dependence and cooling field dependence of exchange-biasing field. Figure 4(a) and Supplementary Fig. 12 show the *M(H)* data measured at various temperatures. For clarity, each magnetization data *M(H)* is shifted along the magnetization axis by a certain value. Prior to each *M(H)* measurement, the sample was warmed up to 340 K and then cooled down to the measurement temperature with an applied field of 0.5 T. It is clearly seen that the shift of the center of hysteresis loops away from the origin of the field-axis. i.e., the exchange-biasing field ($\mu_0 H_{EB}$), emerges below 200 K and gradually increases as the temperature decreases, as shown by the blue symbols presented in Fig. 4(c). Figure 4(b) shows the *M(H)* data measured at *T* = 100 K after the sample was cooled down from 340 K with different cooling fields ($H_{FC}$). $H_{EB}$ increases with $H_{FC}$ and nearly saturates above 0.6 T, as illustrated in Fig. 4(d). And as shown in Fig. 4 (e), $H_{EB}$ switches the sign when the sign of $H_{FC}$ reverses. In addition, we also examined the training effect of the exchange-bias behavior in TbMn$_6$Sn$_6$. The sample was cooled down from 340 K to 100 K with different cooling fields. After each field cooling, complete $M(H)$ loops were then acquired continuously for three times without subsequent warming up and cooling down processes. The obtained $M(H)$ data are shown in Fig. 4(f). Clearly, the sample exhibits prominent training effect with the exchange-biasing field $H_{EB}$ value deceasing dramatically upon repeating the measurements after the first loop cycle, a feature similar to as observed in heterostructures composed of spin glass and ferromagnet bilayers [45]. Note that the transverse transport data shown in Fig. 2(a-c) and the $M(H)$ data shown in Supplementary Fig. 11 were taken after the sample was initially cooled down to 100 K with 1.5 T magnetic field but without the subsequent warming up and cooling down processes for the measurements at each temperature. Thus, the resultant $\mu_0 H_{EB}$ is smaller compared to the values shown in Fig. 4(a) due to the training effect. All these phenomena presented in Fig. 4(b) are characteristic of the exchange-bias behavior, affirming our finding of exchange-bias feature in single crystalline TbMn$_6$Sn$_6$.

**Discussion**

The observation of exchange-bias feature in both magnetization and transverse transport measurements in TbMn$_6$Sn$_6$ is interesting and perplexing. It is worth noting that the dynamic depolarization rate of the μSR signal in TbMn$_6$Sn$_6$ increases appreciably below 200 K as well, indicating the slowing down of magnetic fluctuations [39]. This implies its close correlation with the observed exchange-bias feature. We have performed DFT calculations on the exchange interactions in the Mn Kagome lattice and found that ferromagnetic and antiferromagnetic interlayer exchange couplings between two nearest-neighboring Mn kagome planes compete, compared to the dominant intralayer ferromagnetic coupling (see Supplementary Fig. 13). In addition, it is known that the interlayer couplings between further neighboring Mn Kagome planes in RMn$_6$Sn$_6$, which are sensitive to the R element, also compete to determine their overall spin structures [33]. We speculate that in TbMn$_6$Sn$_6$, as the magnetic fluctuations slows down, the competition of interlayer couplings between Mn spins, together with the antiferromagnetic Tb-Mn interlayer coupling, may lead to the formation of the cluster spin-glass state with antiferromagnetic domains embedded in and coexisting with the bulk ferromagnetic phase of Mn spins. As a result, the interface between antiferromagnetic domains and ferromagnetic phase of Mn spins gives rise to exchange-bias features with asymmetry in the AHE and ANE hysteresis. One may engineer the exchange-bias by partial substitution on the rare earth site without affecting the Mn site to tune the interlayer couplings of Mn spins. The existence of a cluster spin-glass phase in TbMn$_6$Sn$_6$ is evidenced by the ac susceptibility measurements shown in Supplementary Fig. 14. The exchange-bias feature between spin-glass and ferromagnet phases was previously discovered, but in the heterostructure form [45,46].

Finally, we would like to comment on the distinctions and similarities between TbMn$_6$Sn$_6$ and other magnetic topological materials that show ANE [14,22,23,40,47-49]. First, it is worth emphasizing that TbMn$_6$Sn$_6$ exhibits large Berry curvature mainly due to massive Dirac bands near K points. The K-point Dirac gap originates from gapping a vertical nodal line along the K-H direction [see Figure 3(a)]. There is another hot spot of Berry curvature around $k_z = 0.3$, which reveals rich topology in the band structure. In contrast, many other magnetic topological materials show Berry curvature from Weyl points [e.g. Co$_3$Sn$_2$S$_2$ and Mn$_3$(Sn,Ge)] [22,23] or gapped giant nodal rings [e.g., Co$_2$Mn(Ga,Al), Fe$_3$(Ga,Al)] [14,15,48]. The Wiedemann-Franz law is well preserved up to the room temperature for TbMn$_6$Sn$_6$, different from magnetic Weyl semimetals Co$_3$Sn$_2$S$_2$ and Mn$_3$Ge where the Lorenz ratio deviates above 100 K [41,50]. This is because that massive Dirac bands exhibit smoother Berry curvature variation with respect to the Fermi energy, compared to Weyl points. Second, in Figure 5 we show a scatter plot of $S_{xy}$ versus $M$ for different topological magnetic materials. The pink shaded region represents the lower and upper bounds of $S_{xy}$ and $M$ for conventional ferromagnets in which $S_{xy}$ tends to be proportional to $M$ [22]. Note that both Mn$_3$Sn and Mn$_3$Ge are antiferromagnets with very weak canted ferromagnetic moment, and the ANE signal is determined by the Berry curvature of electron bands and thus $S_{xy}$ does not scale linearly with $M$ [22,47]. Interestingly, similar to other topological ferromagnets listed, the $S_{xy}(M)$ of TbMn$_6$Sn$_6$ also falls outside the pink shaded region, implying the contribution of Berry curvature to its ANE signal as supported by the theoretical calculation presented in Fig. 3(c). In addition, while $S_{xy}$ of TbMn$_6$Sn$_6$ is comparable to or slightly smaller than those of Fe$_3$Sn$_2$ [49], Fe$_3$Al [48], Fe$_3$Ga [48], and Co$_2$MnGa [14,40], it has much lower magnetization compared to others. This is an important aspect, particularly considering that TbMn$_6$Sn$_6$ exhibits the unique exchange-bias feature described above. As shown in Fig. 4(c), $M$ of TbMn$_6$Sn$_6$ is even much smaller than exchange-biasing field $\mu_0 H_{EB}$. Thus, as discussed previously in the introduction, one can utilize the large exchange-biasing field to circumvent the stray field interference arising from neighboring thermoelectric modules that are connected in series. This allows for much larger integration density of thermopile and improves the device performance with a significantly enhanced Nernst signal.

In summary, we have reported the prominent AHE, ANE and ATHE behavior of TbMn$_6$Sn$_6$ which is ferrimagnetic Kagome metal hosting a Chern gap near the Fermi level. We show that these anomalous transverse conductivities are associated with large Berry phase in the reciprocal space. Furthermore, we find that TbMn$_6$Sn$_6$ exhibits an exchange-bias feature in both magnetization and transverse conductivity measurements. This, combined with the large ANE, places TbMn$_6$Sn$_6$ as a promising system for transverse thermoelectric devices based on the Nernst effect.

**Methods**

TbMn$_6$Sn$_6$ single crystals were grown using the flux method [37]. It crystalizes in the hexagonal structure with a space group P6/mmm (No. 191) with lattice constants a = b = 5.522 Å, c = 9.004 Å and crystalline angles $\alpha=\beta=90°$, $\gamma=120°$. Magnetic susceptibility measurements of TbMn$_6$Sn$_6$ were carried out using a Superconducting Quantum Interference Device (SQUID) magnetometer. Resistivity and Hall effect measurements were conducted using a Physical Property Measurement System (PPMS). Thermal and thermoelectric transport measurements were performed using a homemade sample puck designed to be compatible with the PPMS cryostat. Type-E thermocouples (Chromel-Constantan) were used for temperature difference measurement. The thermoelectric voltage was measured using K2182A Nanovoltmeters (Keithley). The 'cold-end' of the sample is attached to a piece of oxygen-free high conductivity copper using silver epoxy. A resistive heater (~1 kΩ) was attached to the other end of the sample and the heat current $J_Q$ was applied in the ab plane. The magnetic field was applied along the out-of-plane direction, i.e., c-axis. Both the experimental set-up and data processing of the measured transverse coefficients are described in detail in the Supplementary Information.

**Data availability**

The data that supports the plots within this paper and other findings of this study are available from the corresponding author upon reasonable request.

**Code availability**

The theory calculations are performed via standard density-functional theory package, Vienna Ab initio Simulation Package (VASP, https://www.vasp.at/). The codes that support the plots and data analysis within this paper are available from the corresponding author upon reasonable request.

**Acknowledgements**

H.Z., M.S., and X.K. acknowledge the financial support by the U.S. Department of Energy, Office of Science, Office of Basic Energy Sciences, Materials Sciences and Engineering Division under DE-SC0019259. C.X. is partially supported by the Start-up funds at Michigan State University. B.Y. acknowledges the financial support by the European Research Council (ERC Consolidator Grant ``NonlinearTopo'', No. 815869) and the ISF - Personal Research Grant (No. 2932/21).

**Authors contributions**

X.K. supervised the project. C.X. grew the samples, H.Z., C.X, and M.S. fabricated the devices and performed the measurements. H. Z. analyzed the data. J.K. performed theoretical calculations with support from B.Y.  H.Z. and X.K. wrote the manuscript. All authors commented on the manuscript.

**Competing interests**

The authors declare no competing financial interests.

**Additional information**

**Supplementary information** is available for this paper at (http link).

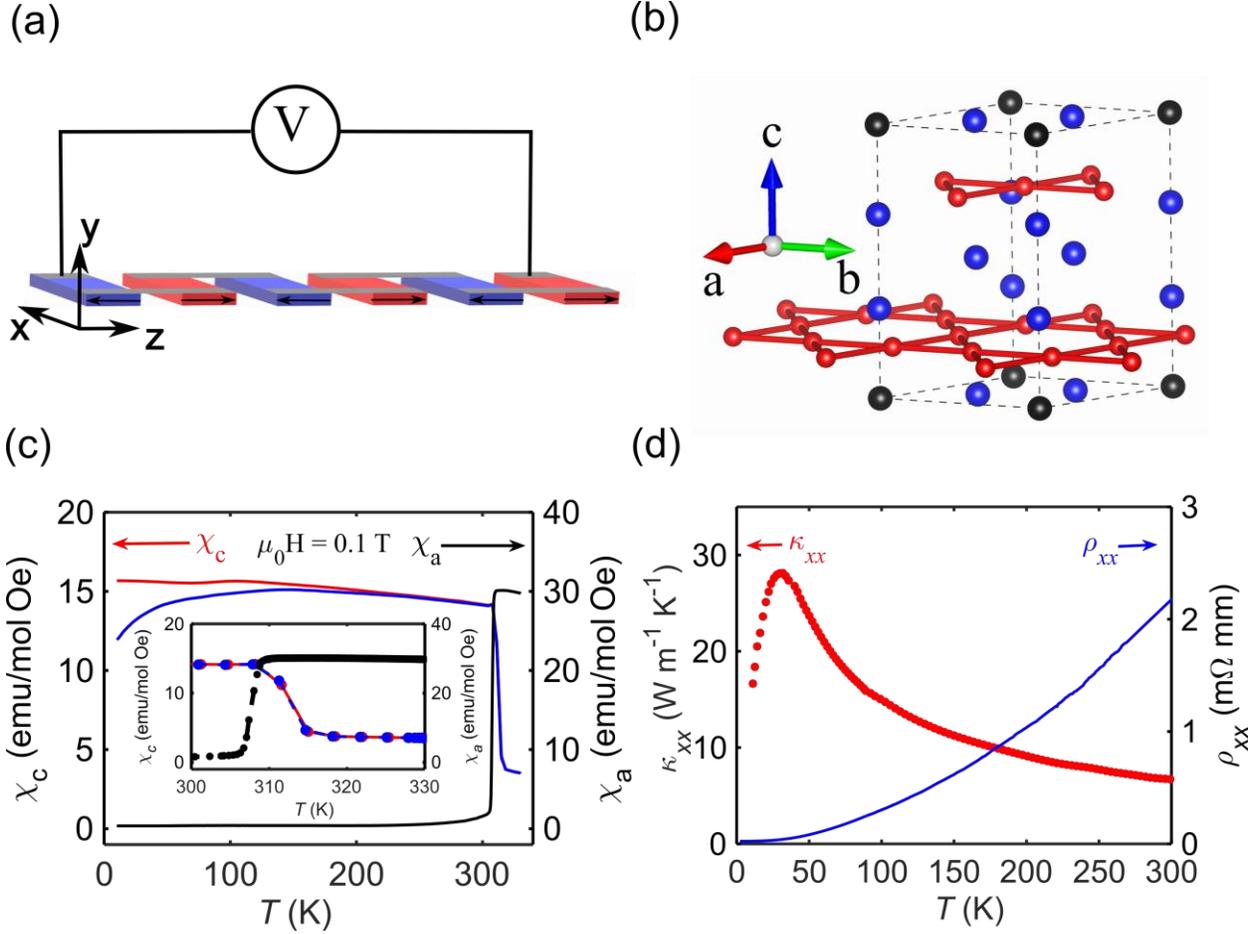

**Figure 1. Crystal structure, magnetic and longitudinal transport properties of TbMn$_6$Sn$_6$.** (a) An illustration of thermoelectric devices utilizing anomalous Nernst effect. The color is an indication of magnetization direction for each module (along the z or -z direction), the gray strips represent electric connections, and the temperature gradient is along the y direction. (b) Schematic crystal structure of TbMn$_6$Sn$_6$ with manganese atoms shown in red, terbium in black and tin in blue. (c) Magnetic susceptibility data for TbMn$_6$Sn$_6$. Blue and red curves are the zero-field cooled and field cooled $\chi_c$ data and black curve is the field cooled $\chi_a$. The applied magnetic field is 0.1 T. (d) Temperature dependence of the longitudinal thermal conductivity $\kappa_{xx}$ and resistivity $\rho_{xx}$.

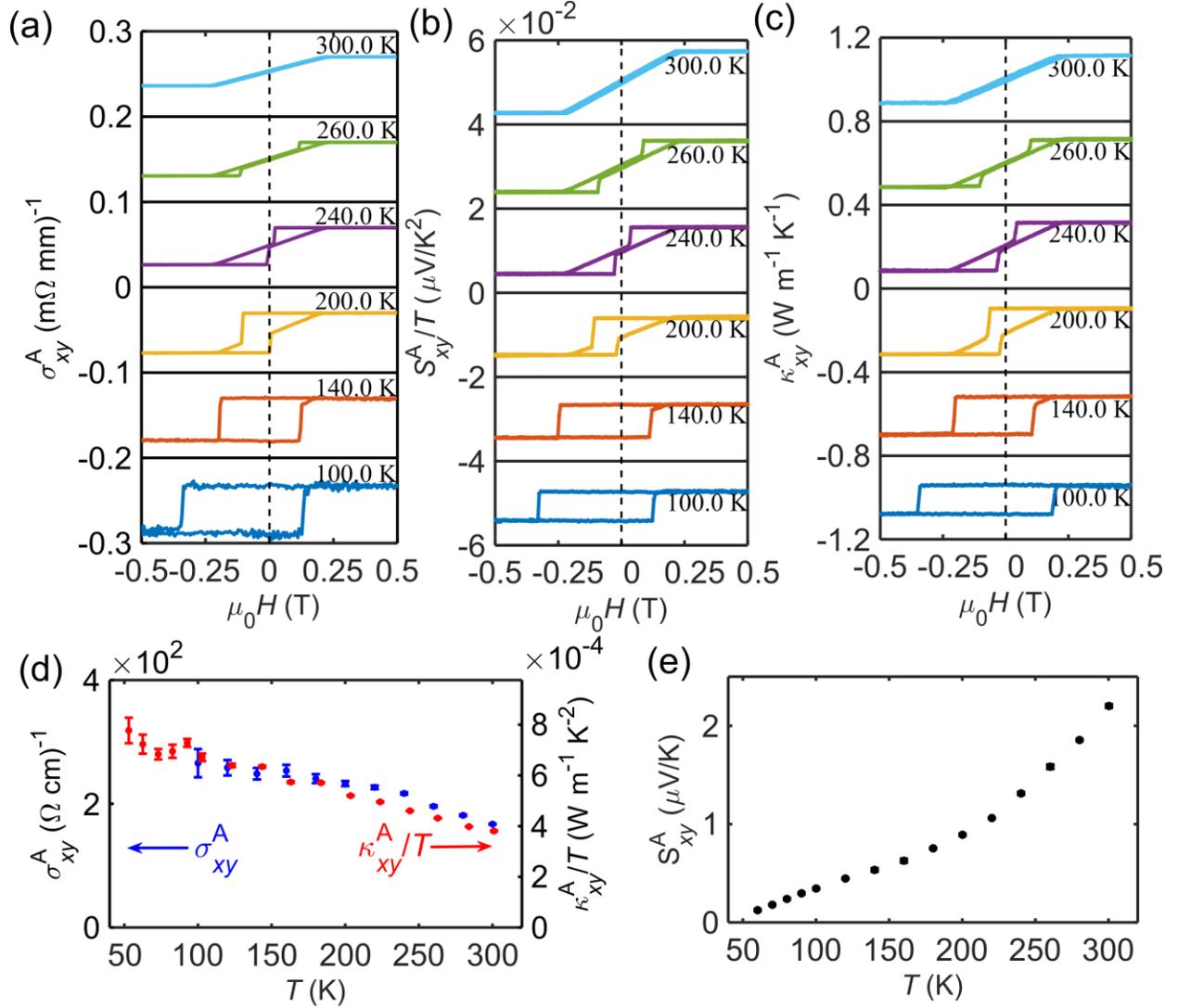

**Figure 2. Hysteresis and temperature dependence of transverse transport properties of TbMn$_6$Sn$_6$.** Magnetic field dependence of anomalous Hall conductivity $\sigma_{xy}^A$ (a), anomalous Nernst coefficients scaled by temperature $S_{xy}^A/T$ (b), and anomalous thermal Hall conductivity $\kappa_{xy}^A$ (c) measured at selected temperatures. (d) Temperature dependence of anomalous Hall conductivity $\sigma_{xy}^A$ and the thermal Hall conductivity normalized to temperature $\kappa_{xy}^A/T$. Error bars were obtained through the fitting methods described in the Supplementary Information. (e) Temperature dependence of anomalous Nernst coefficient ($S_{xy}^A$).

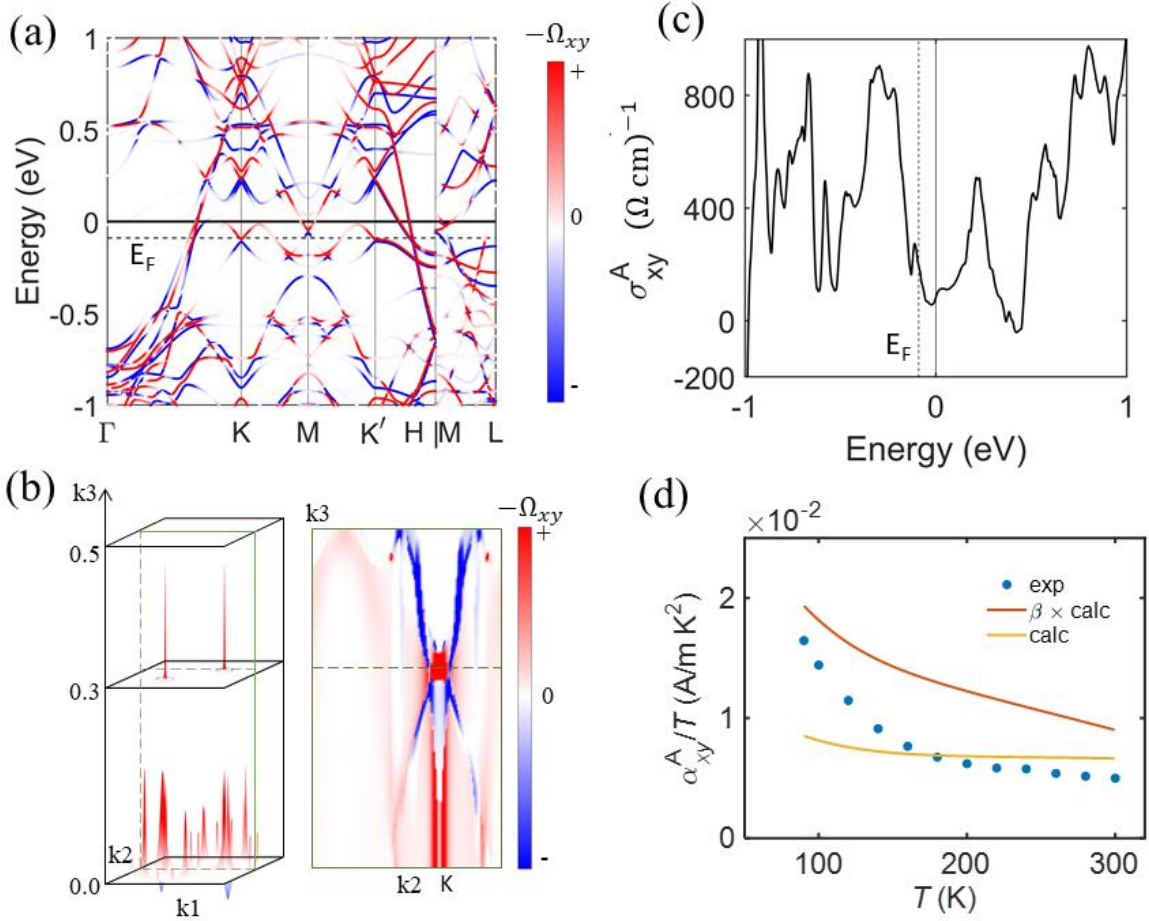

**Figure 3. Electronic structure calculation.** (a) Left panel: electronic band structure and the projected Berry phase ($-\Omega_{xy}$) for bands near the Fermi energy. The experimental Fermi energy ($E_F$, dashed line) is 90 meV lower than the charge-neutral point (zero). (b) Visualization of Berry curvature distribution in 3D Brillouin zone at $E_F$. (c) Calculated intrinsic anomalous Hall conductivity ($\sigma_{xy}^A$) as a function of the Fermi energy at zero temperature. (d) Temperature dependence of the measured (blue), the calculated (yellow), and the scaled calculated (orange) transverse thermoelectric conductivity normalized by temperature $\alpha_{xy}^A/T$.

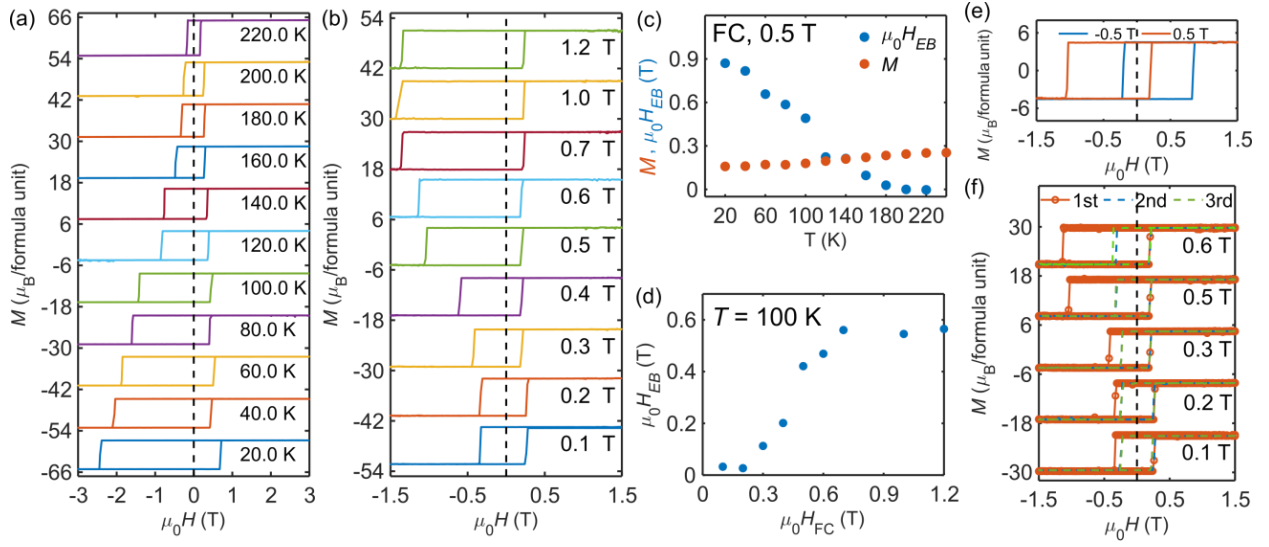

**Figure 4. Exchange-bias behavior of TbMn$_6$Sn$_6$.** (a) Magnetization hysteresis loops measured at various temperatures after the sample was cooled down from 340 K to the measurement temperature with 0.5 T magnetic field prior to each measurement. (b) Magnetization hysteresis loops measured at 100 K after the sample was cooled down from 340 K to 100 K with different magnetic field prior to measuring each loop. (c) Temperature dependence of the exchange-biasing field $\mu_0 H_{EB}$ extracted from Figure (a). Orange dots represent the saturated magnetizations at respective temperatures. (d) The cooling field ($H_{FC}$) dependence of the exchange-biasing field extracted from Figure (b). (e) Magnetization hysteresis loops measured at $T = 100$ K with 0.5 T and -0.5 T cooling field. (f) Magnetization hysteresis loops measured at $T = 100$ K upon repeating the measurements after the sample was cooled down with various cooling fields.

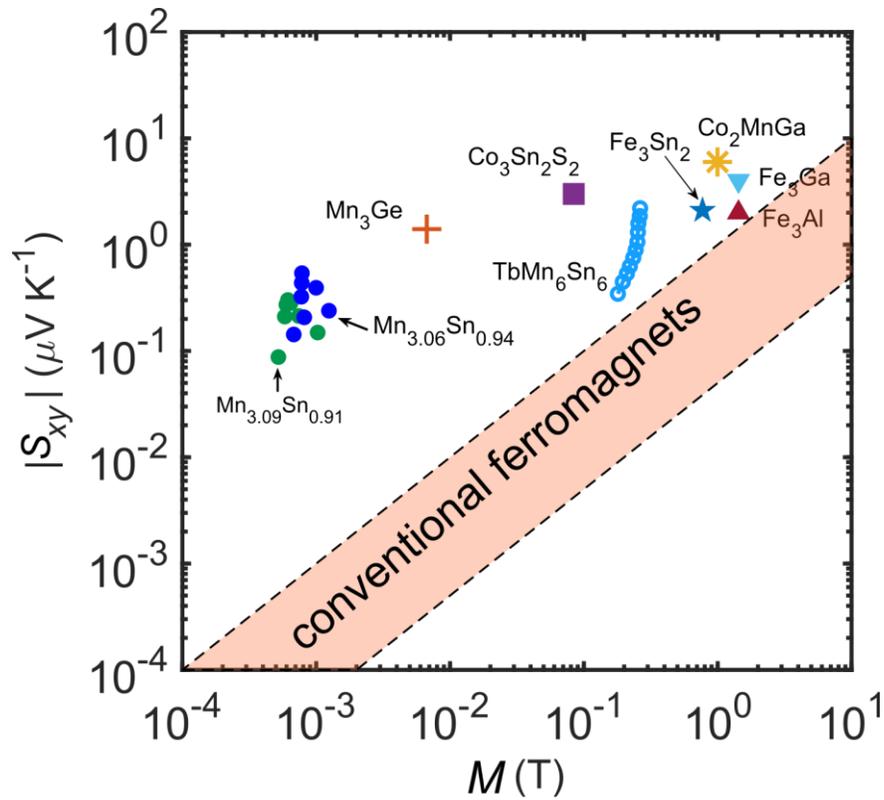

**Figure 5. A comparison of anomalous Nernst signal between TbMn₆Sn₆ and other recently discovered topological metals.** Data points of materials other than TbMn$_6$Sn$_6$ are adapted from previous studies [22,23,40,47-49]. The pink shaded region represents the lower and upper bounds of $S_{xy}$ and $M$ for conventional ferromagnets [22].